\documentclass[prd,twocolumn,fleqn,showpacs,showkeys,amsmath]{revtex4}

\usepackage{amsfonts}
\usepackage{epsf}
\usepackage{graphicx}
\usepackage{psfrag}
\usepackage{amssymb}
\usepackage{color}
\usepackage{amsmath}
\usepackage{chemarr}

\newcommand{\rme}{\mathrm{e}}
\newcommand{\rmi}{\mathrm{i}}
\newcommand{\rmd}{\mathrm{d}}

\newcommand{\bfr}{\mathbf{r}}

\renewcommand{\qquad}{\hspace*{25pt}}

\begin{document}

\setcounter{page}{1}%

\title{Formation of Dimers in Axion-Like Dark Matter Using the Feshbach Resonance}

\author{A.M.~Gavrilik\footnote{e-mail: omgavr@bitp.kiev.ua}, A.V.~Nazarenko\footnote{e-mail: nazarenko@bitp.kiev.ua}}
\affiliation{Bogolyubov Institute for Theoretical Physics of NAS of Ukraine, \\ 
14b, Metrolohichna Str., Kyiv 03143, Ukraine}%


\begin{abstract}
Within the model of self-gravitating Bose--Einstein condensate (BEC) dark matter (DM)
it is argued that the axion-like self-interaction of ultralight bosons provides
the existence of rarefied and dense phases, which are predicted earlier on
the base of the models with polynomial-like self-interactions. Associating the very
short  scattering length in BEC DM with the predominant participating composites
of few DM particles, we attempt to form a dimer of two particles at a quantum
mechanical level, using a smooth $\mu$-deformation of the axion cosine-like
potential and replacing the field-dependent argument with the distance between
particles. Part of the obtained results concerns potential two-particle scattering
with $\mu$-deformed interaction, and they allow us to focus on a special option
with unique values of the deformation parameter $\mu=1$ and the coupling constant.
In this case of the potential with an infinite scattering length, we get a rather
simple solution for the dimer in the ground state. We involve two-channel
scattering and Feshbach resonance to describe the formation of a dimer in space.
Specifying the parameters of interactions, we reveal a long-lived resonance that
occurs when a pair of particles jumps between the open and closed scattering
channels with close energy values. This indicates the possibility of participation
of such dimers in forming BEC DM halo of galaxies.
\end{abstract}



\maketitle

\section{Introduction}

Axion-like bosons belong to most popular candidates of the Bose--Einstein
condensate (BEC) dark matter (DM), the properties of which are studied in
many theoretical works~\cite{PWW,Sin,Lee,Hu,Sahni,Ferreira,Bohmer,SY09,
Harko2011,axion3,axion4,axion5,Chavanis2}.
Their being involved in the description of astrophysical phenomena should
borrow the chiral cosine-like self-interaction~\cite{PQ77,Wit80}, and also
requires engaging the gravity (usually treated
separately of the unification theory), the account of which leads to breaking
the inherent symmetry. Note that there are also a large number of works,
including experimental ones~\cite{Brad03}, that study the interaction of
axions with other substances and their transformation (due to
the Primakoff effect, see~\cite{Raff}), as well as their role in cosmology~\cite{SY09}.

The condensate properties of the axion-like DM, taking into account
both the leading pairwise contributions to the
self-action~\cite{Bohmer,Harko2011,axion5} and the next
three-particle corrections~\cite{Chavanis2,GKN20,GN21}, are
promising for further consideration and application of axion-like
particles. Having got a number of characteristics, the dilute and
dense phases along with the phase transitions are revealed in
BEC~\cite{Chavanis2,GKN20,GN21}. Besides, analysis of these effects
and ways to better describe the observables suggests the existence
of molecule-like composites~\cite{GN22} and the relevance    
of deformation-based description~\cite{muBose1,muBose2,Nazar}.
Theoretically, these possibilities are considered as very important,
when dealing with the dark sector.

Moreover, there are predictions concerning both the first-order phase transition
at zero temperature with changing the interaction parameters and its
influence on the rotation curves of the DM halo of dwarf galaxies~\cite{GN21}. 
  Physically, this is associated with quantum fluctuations in regions with
a relatively high number density of ultralight particles, where the
three-particle effects become significant. There are indications
that quantum entanglement may be involved~\cite{Qent,GN21}. At first
glance, composites would be also produced under these conditions as
well. However, as was shown earlier \cite{GN22}, the appearance of
particle complexes (``molecules'') is not supported at high density
because of disintegration stimulated by frequent collisions, but may
be caused by a large scattering length of particle interaction. It
is clear that this information is able to shed light on the
characteristics and behavior of initial DM particles, the nature of
which have not yet been identified.

In general, it is natural to assume that the DM consists of
particles of different sorts, including composites. What we observe
and describe now is mainly the result of self-interaction, that is,
a steady state with a vanishingly small scattering length, confirmed
by numerous models based on the Gross--Pitaevskii
equation~\cite{Harko2011,Chavanis2,GN21,GN22}. Therefore, we need to
explain these peculiarities which can be associated with the
particle states separated by a large energy gap.

Heuristically, formation of the simplest molecules of two and three
particles is explored in Ref.~\cite{GN22}. Here we have in mind some
analogy with scattering processes in BEC studied in the
laboratory~\cite{Grimm10}. Assuming this to be admissible, we appeal
to the quantum mechanical formation of a molecule (dimer) of two
particles, borrowing the ideas of the Feshbach resonance and using
two scattering channels~\cite{Fesh62,Joa75,Yam93,Grimm10}. Although it
is appropriate to include auxiliary influences in our consideration,
a more detailed analysis of which is an independent task, our goal
is to disclose a plausible mechanism for the formation of bound
states during the Universe evolution. Thereby, we emphasize the fact
that one good potential is not sufficient to form DM molecules in
space.

Anyway, the choice of the self-interaction potential is decisive.
Having gained an idea of the nontrivial phase structure of DM with two-
and three-particle interactions and its manifestations in observables~\cite{GKN20,GN21},
we want to show here that the model with the cosine-like interaction
mentioned above should also lead to similar consequences.
Obviously, the already used self-interactions of the polynomial form
become treated as the expansion terms of the total potential.
It is important that the cosine-like generalization not only complicates the form
of interaction, but also reduces the number of independent parameters.
We focus on different phases of dark matter with axion-like interaction
in the spherically symmetric case, when the main function we find
is the spatial distribution of particles in BEC.

Further, what concerns the problem of the formation of dimers in DM,
we follow our strategy, according to which the structured
(composite) particles are described by deformed
functions~\cite{GM1,GM2,GM3}. The approach based on
$\mu$-deformation has already been successfully tested in our
works~\cite{muBose1,muBose2,Nazar}. Here, we replace the ordinary
cosine potential of axion-like bosons with the $\mu$-deformed cosine
introduced earlier~\cite{muBose2}. Of course, when $\mu=0$, the
usual form of the cosine function is restored. The physical
attractivity of such a modification lies in the possibility of
smooth transforming the periodic function into a typical pairwise
potential, the form of which is often encountered in classical and
quantum mechanics.

Despite the violation of the symmetry of the corresponding field
theory, the expediency of this deformation can be justified by its
efficiency. This indeed follows by developing the quantum mechanical
picture of two-particle scattering, in which the distance between
particles is used as a variable for the potential instead of the
axion field. Devoting a part of our study to the scattering
properties, when the deformation parameter $\mu$ strongly affects
the scattering length, we discover the possibility of the dimer
formation from a pair of DM particles. Namely, we use the deformed
potential with suitable value $\mu=1$ in the two-channel resonance
scattering to describe the creation of a single molecule, leaving
aside the many-particle condensate in the present treatment.

The paper is organized as follows. In Sec.~\ref{S2} we show the
existence of two phases of BEC DM on the base of a pair of different
initial conditions for the stationary Gross--Pitaevskii equation
with cosine-like and gravitational interactions at fixed values of
the coupling constants and chemical potential. The $\mu$-deformed
self-interaction is introduced in Sec.~\ref{S3} instead of the
typical cosine-like one for axions to account for the structured
(composite) particles of DM. The scattering properties of such
potentials, which we treat as the functions of distance between
particles, are studied in the Born approximation for $s$-waves. In
the case of $\mu=1$, we find an exact solution to the problem with
attractive potential which corresponds to a two-particle molecule (dimer)
of DM particles. To describe the dimer formation, we appeal to
the resonance scattering in Sec.~\ref{S4}, where the two-channel
mechanism is applied in accordance with Feshbach. This allows to
analyze the conditions of forming DM composites in the Universe. The
final Section~\ref{S5} is devoted to discussion of implications and
concluding remarks.

\section{\label{S2}Gross--Pitaevskii Equation for Axion-Like DM and Its Solution}

We start with formulating stationary macroscopic model of
gravitating Bose--Einstein condensate (BEC) of ultralight bosons
with axion-like interaction, restricting ourselves to the spherical
symmetry and the absence of hydrodynamic flows. Let the BEC with a
constant chemical potential $\tilde\mu$ be described by real
function $\psi(r)$ of radial variable $r=|\bfr|$, and $\psi^2(r)$
defineing a local particle density. The development of $\psi(r)$ in
the ball $B=\{\bfr\in\mathbb{R}^3|\, |\bfr |\leq R\}$ follows from
both vanishing variation of the energy functional $\Gamma$  and the
Poisson equation for the gravitational potential $V(r)$:
\begin{eqnarray}
&&\Gamma=4\pi\int_0^R\left\{\frac{\hbar^2}{2m}(\partial_r\psi)^2
-\tilde\mu\psi^2+mV\psi^2\right.\nonumber\\
&&\hspace*{6mm}
\left.+\frac{U}{v}\left[1-\cos{\left(\sqrt{v}\psi\right)}\right]
-\frac{U}{2}\psi^2\right\}\,r^2\,\rmd r;\label{G1}\\
&&\Delta_r V=4\pi Gm\psi^2,
\end{eqnarray}
where the radial part of Laplace operator $\Delta_r$ and its inverse $\Delta^{-1}_r$
of variable $r$ are
\begin{eqnarray}
&&\Delta_rf(r)=\partial^2_rf(r)+\frac{2}{r}\,\partial_rf(r),\label{Dlt}\\
&&\Delta^{-1}_rf(r)=-\frac{1}{r}\int_0^rf(s)\,s^2\,\rmd s-\int_r^{R}f(s)\,s\,\rmd s;
\label{InvDelta}
\end{eqnarray}
$R$ is the radius of the ball, where the matter is located.

The axion-like cosine-interaction is characterized by two constants $U$ and $v$,
which have the dimensions of energy and volume, respectively. In the particle
physics, they are related with the axion mass and decaying constant
$f_{\mathrm{a}}$ as $U=mc^2$, $v=\hbar^3c/(mf^2_{\mathrm{a}})$. Note their
relativistic nature, which is exploited in \cite{Chavanis2}.

To analyze the general properties of the model, we reformulate it in dimensionless
variables:
\begin{eqnarray}
&&\xi=\frac{\sqrt{mU}}{\hbar}\,r,\quad
\chi(\xi)=\sqrt{v}\,\psi(r),\label{mp1}\\
&&\xi_B=\frac{\sqrt{mU}}{\hbar}\,R,\quad
A=4\pi\frac{G\hbar^2m}{U^2v},\label{mp2}\\
&&\nu=2u+2A\int_0^{\xi_B}\chi^2(\xi)\,\xi\,\rmd\xi+1,\quad u=\frac{\tilde\mu}{U},
\label{mp3}
\end{eqnarray}
where $\nu$ plays the role of effective chemical potential, which
absorbs the constant term of axion interaction and the
gravitational potential at the center $\xi=0$:
\begin{equation}
V_{\mathrm{grav}}(0)=-A\int_0^{\xi_B}\chi^2(\xi)\,\xi\,\rmd\xi.
\end{equation}
In our study, $\nu$ is regarded as a free parameter.

The model equations in terms of the wave-function $\chi(\xi)$ and
auxiliary gravitational potential $\Phi(\xi)$ read
\begin{eqnarray}
&&\left(\Delta_{\xi}+\nu\right)\,\chi-2A\Phi\chi-\sin{\chi}=0,\label{feq1}\\
&&\Phi(\xi)=-\frac{1}{\xi}\int_0^\xi \chi^2(s)\,s^2\,\rmd s+\int_0^{\xi}\chi^2(s)\,s\,\rmd s,
\label{feq2}
\end{eqnarray}
where $\Delta_\xi\Phi(\xi)=\chi^2(\xi)$ is satisfied, and $\Phi(0)=0$.

To obtain a finite and stable solution $\chi(\xi)$, the non-linear
Eqs.~(\ref{feq1})-(\ref{feq2})
should be numerically integrated under the following conditions: $\chi(0)<\infty$,
$\chi^\prime(0)=0$, $\chi^{\prime\prime}(0)<0$. For given $A$ and $\nu$,
the finite initial value $\chi(0)=z$ should be positive solution
of the equation:
\begin{equation}\label{icon}
2Az^2+\left(\nu-\frac{\sin{z}}{z}\right)(\nu-\cos{z})=0,
\end{equation}
which is derived by substituting $\chi(\xi)=\chi(0)+\chi^{\prime\prime}(0)\xi^2/2$
into (\ref{feq1})-(\ref{feq2}) at $\xi\to0$, and by determining first
$\chi^{\prime\prime}(0)$.
Note that the requirement $\chi^{\prime\prime}(0)<0$ is equivalent to
imposing $\nu>\sin{z}/z$.

The absence of a solution $z$ at given pair $(A,\nu)$ means that
$\chi(\xi)=0$ everywhere. It happens at $\nu>\nu_{\mathrm{max}}$,
where $\nu_{\mathrm{max}}(A)$ is also found numerically from
(\ref{icon}). For $\nu<\nu_{\mathrm{max}}$, two branches of
$\chi_0(\nu)$ can occur, which witness the possibility of two
regimes existence and a first-order phase transition in the model.

\begin{figure}[htbp]
\includegraphics[width=8cm,angle=0]{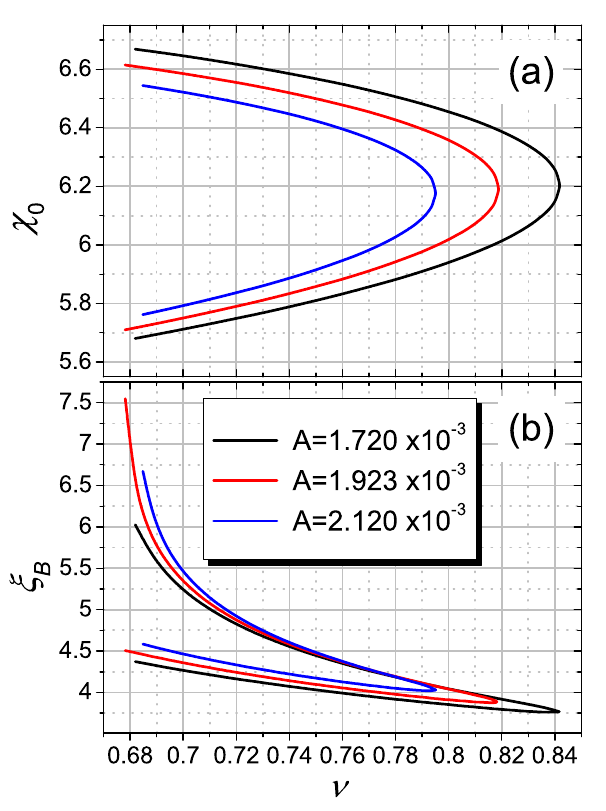}
\caption{\label{chi0}The initial value $\chi_0=\chi(0)$ (a) and the first
zero $\xi_B$ (b) of $\chi(\xi)$ versus parameter $\nu$ for different $A$.
The presented curves are limited by
$\nu^{\mathrm{black}}_{\mathrm{max}}=0.841954176$,
$\nu^{\mathrm{red}}_{\mathrm{max}}=0.818967072$,
$\nu^{\mathrm{blue}}_{\mathrm{max}}=0.795228679$.
At $\nu>\nu_{\mathrm{max}}$ one has $\chi_0=0$ and $\chi(\xi)=0$.}
\end{figure}

\begin{figure}[htbp]
\includegraphics[width=7.8cm,angle=0]{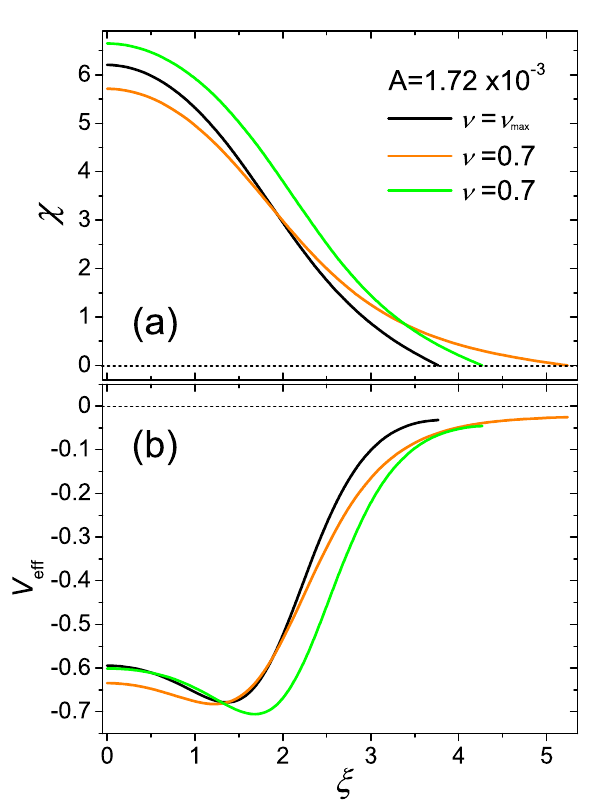}

\vspace*{-1.5mm}
\caption{\label{chi}A particular form of the wave function $\chi(\xi)$ (a) and
the effective potential $V_{\mathrm{eff}}(\xi)$ (b) for the fixed parameters $A$
and $\nu$. Green and orange lines are obtained with the same $A$ and $\nu$,
but differ in the initial values $\chi_0$, see Fig.~\ref{chi0}(a). The value of
$\chi_0$ for green curves belongs to the upper branch of $\chi_0(\nu)$ in
Fig.~\ref{chi0}(a), while the orange lines are plotted using $\chi_0$ of
the lower branch of $\chi_0(\nu)$.}
\end{figure}

Assuming that gravity is weaker than self-interaction, we consider
the parameter $A\sim10^{-3}$. Then, analyzing Eq.~(\ref{icon}), the
two-phase structure of DM appears for $z>\pi$. Indeed, choosing $A$
as in Fig.~\ref{chi0}(a), the upper branch of $\chi_0(\nu)$ is
represented by $\chi_0\in[2\pi;5\pi/2]$, while the lower branch of
$\chi_0(\nu)$ gives us $\chi_0$ in the interval $[3\pi/2;2\pi]$. In
the both cases, one has $\nu>\sin{\chi_0}/\chi_0$.

Although the axion field in QCD is usually assumed to be weak ($\chi_0<\pi$)
near the true vacuum, the situation looks different in a non-relativistic
model of DM with Newtonian interaction. Also, there is no invariance there
under the global transformation $\chi\to\chi+2\pi$, violated by gravitation.
We associate such a discrepancy with the consideration of the condensate
in a finite volume (galactic halo of DM), and we admit its correction when
describing gravity as the geometry of space-time.

Indeed, Fig.~\ref{chi0}(b) demonstrates the value of (first) zero $\xi_B$ of function
$\chi(\xi)$ (that is $\chi(\xi_B)=0$), which limits the system size in our model.
This is found by integrating the Eqs.~(\ref{feq1})--(\ref{feq2}) for given
$A$ and $\nu$.

The typical solutions for the wave function $\chi(\xi)$ are presented in
Fig.~\ref{chi}(a), where $\xi\in[0;\xi_B]$. They describe a core with a finite
value of the particle (and mass) density in the center $\xi=0$. Besides, an additional information
may be extracted from an effective potential $V_{\mathrm{eff}}$ (see Fig.~\ref{chi}(b))
as a function of radial variable $\xi$:
\begin{eqnarray}
&&V_{\mathrm{eff}}(\xi)=V_{\mathrm{grav}}(0)+A\Phi(\xi)
+\frac{1}{2}\left(\frac{\sin{\chi(\xi)}}{\chi(\xi)}-1\right),\\
&&\left[-\frac{1}{2}\Delta_\xi+V_{\mathrm{eff}}(\xi)\right]\chi(\xi)=u\chi(\xi),
\end{eqnarray}
where Eq.~(\ref{feq1}) is re-written in the form of Schr\"odinger equation
with the chemical potential $u$ from (\ref{mp3}).

Fig.~\ref{chi}(b) indicates the presence of a minimum of the
attractive potentials colored in green and orange. Since these
minima have different depths for the same $A$ and $\nu$, two
macroscopic phases become possible in the system. The being of the
particles in one of the two phases is conditioned by the applied
factors e.g. pressure~\cite{GKN20,GN21}.

Outside the system at $\xi>\xi_B$, where there is no the matter, $V_{\mathrm{eff}}(\xi)$
is continuously extended by the gravitational potential $-A{\cal N}/\xi$, where
${\cal N}$ is the total number of particles in the ball $\xi\leq\xi_B$.

To estimate the characteristics of our model, we relate them with
the dimensionless parameters (\ref{mp1})-(\ref{mp3}). Following the
model from Ref.~\cite{GKN20}, which allows to separate approaches to
describing the core and tail of the DM halo of (dwarf) galaxies due
to the different role of self-interaction, we intend to estimate the
size scale $r_0$ and the central mass density $\rho_0$ of the core,
so that $r=r_0\xi$ and $\rho(\xi)=\rho_0\chi^2(\xi)/\chi^2_0$. At
the same time, we need to control the parameters $U$ and $v$ of
axion-like interaction.

As it was stated in Ref.~\cite{Chavanis2}, the models involving
axions for describing compact objects (such as axion stars) and for
cosmology lead to different parametrizations to be taken into
account. First of all, this concerns the different mass ranges.
While cosmological models constrain the axion mass as
$10^{-7}\,\text{eV}<mc^2<10^{-2}\,\text{eV}$, ultralight particle
models suggest that $mc^2\sim10^{-22}\ \text{eV}$, typical for the
fuzzy DM. Physically, the choice of a smaller particle mass affects
the formation of certain structures in the Universe~\cite{FMT08},
which we are trying to account for. It stimulates us to fix
$m\sim10^{-22}\ \text{eV}\,c^{-2}$. Although the parameter $U$ must
be proportional to $mc^2$ to ensure transmutation, we have no reason
to declare identity in the non-relativistic case. We only admit that
the value of $U$ at relatively large magnitude of the axion field
allows the dominance of repulsion, caused by self-interaction, over
gravitation.

Relating the volume $v=\hbar^3c/(mf^2_{\mathrm{a}})$ with the axion
decay constant $f_{\mathrm{a}}$, we note that
$10^{18}\,\text{eV}<f_{\mathrm{a}}<10^{21}\,\text{eV}$ in cosmology.
On the other hand, its value may be larger in the models with
ultralight particles~\cite{Chavanis2}.

In our model, it is easy to estimate that
\begin{eqnarray}
f_{\mathrm{a}}&\simeq&3.304\times10^{19}\,\text{eV}\,
\left[\frac{\rho_0}{10^{-19}\,\text{kg}\,\text{m}^{-3}}\right]^{1/2}
\nonumber\\
&&\times\left[\frac{mc^2}{10^{-22}\,\text{eV}}\right]^{-1}
\left[\frac{\chi_0}{2\pi}\right]^{-1}
\end{eqnarray}
for $\rho_0\sim10^{-19}\,\text{kg}\,\text{m}^{-3}$ while a mean mass density
is usually assumed to be of the order $10^{-20}\,\text{kg}\,\text{m}^{-3}$.

The characteristic scale is evaluated to be
\begin{eqnarray}
r_0&\simeq&0.138\,\text{kpc}\,\left[\frac{\rho_0}{10^{-19}\,\text{kg}\,\text{m}^{-3}}\right]^{-1/4}
\left[\frac{mc^2}{10^{-22}\,\text{eV}}\right]^{-1/2}
\nonumber\\
&&\times
\left[\frac{A}{2\times10^{-3}}\right]^{1/4} \left[\frac{\chi_0}{2\pi}\right]^{1/2}.
\end{eqnarray}
Indeed, $r_0$ is applicable for measuring the size of the central part
of the DM halo as $R=r_0\xi_B$, but should be fixed together with
the total mass $M$.

To confirm the presence of a first-order phase transition in the
model under consideration, it is needed to develop a statistical
approach~\cite{GKN20,GN21}, which is omitted here. It is expected
that the discontinuous change in density at zero temperature will be
caused by a change in pressure/compression. The effect of different
phases of DM on the observables, as well as on the rotation curves,
is an interesting problem for a separate study.

Contrary to the background of opening prospects, the results of
which can be predicted using the known
models~\cite{Chavanis2,GKN20,GN21}, the cosine self-interaction is
worth to study due to additional possibility of smooth modification,
which suggests a number of new effects in DM.

\section{\label{S3}$\mu$-Deformed Interaction}

Although we left aside the statistical description of the
condensate, it is foreseen there that the scattering length in the
model with cosine-like self-interaction will be vanishingly small
(smaller than the Planck one), as in similar models with
few-particle interactions, even in the Thomas--Fermi
approximation~\cite{Harko2011,Chavanis2,GKN20,GN21}. It can be
assumed that such a steady state is due to the interaction of the
initial DM particles with a long scattering length leading to the
formation of complexes~\cite{GN22}. In addition, various macroscopic
phases imply the ability of DM structuring, that is, its various
fractions. Taking this into account, we deduce that the initial
interaction between DM particles can be more complicated.

An effective way to extend the abilities of the used interaction is
its smooth deformation. Choosing the kind of deformation to achieve
our goals, we rely on existing experience that indicates the
relationship between the so-called $\mu$-deformation in particle
systems and forming composites~\cite{GM3}. Technically, this needs
introducing a numerical parameter $\mu$, while the original form of
interaction is recovered at $\mu=0$, without involving other
physical entities. In this way we intend to introduce
$\cos_{\mu}{x}$ instead of ordinary function $\cos{x}$.

To describe its properties that are important for physics, we first turn to
the quantum mechanical problem of potential scattering, where $\cos_{\mu}{x}$
plays the role of a pair interaction potential depended on the distance $r$
between particles. Then the useful task is to establish a connection between
the scattering characteristics and the deformation parameter~$\mu$. Moreover,
we show that the quantum problem for $\mu=1$ reveals the bound state of
a two-particle molecule as a DM dimer.

\subsection{Definitions}

Using the $\mu$-bracket (see \cite{muBose2} and references therein),
\begin{equation}
[x]_\mu=\frac{x}{1+\mu x},
\end{equation}
we define the $\mu$-deformed trigonometric functions as
\begin{equation}
\sin_{\mu}{x}=\sum\limits_{n=0}^\infty (-1)^n\frac{x^{2n+1}}{[2n+1]_\mu!},\quad
\cos_{\mu}{x}=\sum\limits_{n=0}^\infty (-1)^n\frac{x^{2n}}{[2n]_\mu!},
\end{equation}
where $[n]_\mu!=[1]_\mu\,[2]_\mu\ldots[n]_\mu$.

Contracting the series, we can write
\begin{eqnarray}
&&\hspace*{-4mm}
\sin_{\mu}{x}=I_{\mu}(x)\sin{\varphi_{\mu}(x)},\quad
\cos_{\mu}{x}=I_{\mu}(x)\cos{\varphi_{\mu}(x)},\label{trig1}\\
&&\hspace*{-4mm}
I_{\mu}(x)\equiv(1+\mu^2x^2)^{-\frac{1+\mu}{2\mu}},\quad
\varphi_{\mu}(x)\equiv\frac{1+\mu}{\mu}\arctan{(\mu x)}.
\nonumber
\end{eqnarray}

\begin{figure}
\begin{center}
\includegraphics[width=8cm]{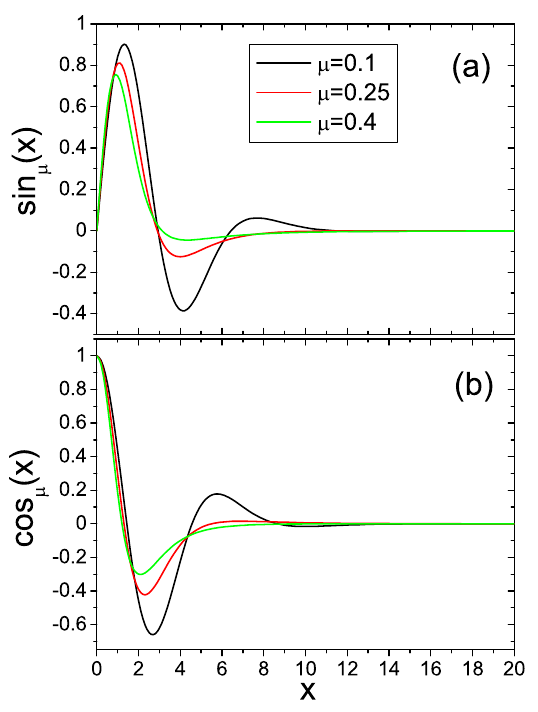}
\end{center}
\vspace*{-8mm}
\caption{Basic $\mu$-deformed trigonometric functions.}\label{fig3}
\end{figure}

Another equivalent form of $\cos_{\mu}{x}$ is also available:
\begin{equation}\label{cosm}
\cos_{\mu}{x}=\frac{1}{2}\left[\left(1+\rmi\mu x\right)^{-1-\frac{1}{\mu}}
+\left(1-\rmi\mu x\right)^{-1-\frac{1}{\mu}}\right].
\end{equation}
Although we are interested in $\mu>0$, note that $\cos_{\mu}{x}=1$
at $\mu=-1/2$ for any $x$.

Behavior of the deformed functions is shown in Fig.~\ref{fig3},
where the $\mu$-deformed cosine in Fig.~\ref{fig3}(b) does imitate
the behavior of the pair interaction potential, associating $x$
with dimensionless distance between particles, and it continuously
changes shape as $\mu$ varies. This fact makes it useful for
studying the properties of particle systems.

Beyond our vision, we get the following asymptotic:
\begin{equation}
\cos_\mu{x}\sim-\sin{\left(\frac{\pi}{2\mu}\right)}\,(\mu x)^{-1-\frac{1}{\mu}},\qquad
x\to\infty,
\end{equation}
where $\mu\not=1/(2n)$ for an integer number $n$. In the case of $\mu=1/(2n)$,
the term of the order $\sim(\mu x)^{-\frac{1}{\mu}}$ becomes significant.

Using the asymptotic, the interaction potentials $x^{-n}$ with $n\geq3$
are traditionally considered as short-range ones, while potentials with
$n\leq2$ belong to long-range ones~\cite{LL}. According this classification,
the function $\cos_\mu{x}$ asymptotically describes short-range potentials
for $\mu<1/2$ and long-range potentials for $\mu>1/2$. In the latter case,
a large scattering length $a$ can be achieved, which is justified below
using the relation with the scattering phase shift $\delta$.

Authors of \cite{Sade} deduced the special properties of phase shift
in $s$-channel near a collision threshold when wave number $k$ vanishes.
If the potential varies as $x^{-n}$ at large $x$, then $\tan{\delta(k)}\sim k$
for $n\geq3$, and $\tan{\delta(k)}\sim k^{n-2}$ for $n\leq3$.
This tells us what happens, when computing the scattering
length $a$ for $n=1+1/\mu$:
\begin{equation}\label{SL}
a=-\lim\limits_{k\to0}\frac{\tan{\delta(k)}}{k}.
\end{equation}

Requiring a unitary regime at $a\to\pm\infty$, we single out
the case $\mu=1$ when $\cos_{\mu}{x}\simeq-x^{-2}$ is similar
to the Efimov hyperspherical potential with interesting scaling
properties~\cite{BrH06}. We pay special attention to this situation
below.

\subsection{Potential Scattering Properties}

Let the interaction potential of two particles,
which depends on the distance $r$ between them, be equal to
\begin{equation}
V_{\mu}(r)=U_0\cos_{\mu}{\frac{r}{r_0}},
\end{equation}
where $U_0=V_\mu(0)>0$ and $r_0$ are constant parameters.

Then, the stationary Schr\"odinger equation for two identical particles
in the center-of-mass reference frame takes on the form:
\begin{equation}
-\frac{\hbar^2}{2m}\Delta\Psi+V_\mu(r)\Psi=E\Psi,
\end{equation}
where $\Delta$ is the Laplace operator in spherical coordinates
$(r,\theta,\varphi)$; $m$ is the reduced mass of two particles.

Focusing on the study of low-energy processes in the spherically
symmetric problem, we consider only the $s$-channel component
$\psi(r)$ of the total wave function
$\Psi(r,\theta,\varphi)=\psi(r)+\Psi_{L>0}(r,\theta,\varphi)$,
discarding the series $\Psi_{L>0}$ in spherical harmonics with
angular momentum $L>0$ along with the corresponding centrifugal
term in equation. This leads to the one-dimensional equation:
\begin{equation}\label{Sch1}
-\frac{\hbar^2}{2m}\Delta_r\psi+V_\mu(r)\psi=E\psi,\qquad
\Delta_r=\frac{1}{r}\frac{\rmd^2}{\rmd r^2}r,
\end{equation}
where $\Delta_r$ is the radial part of Laplace operator.

For convenience, we use dimensionless variables as $x$
(distance), $k$ (wave number), and $g$ (coupling constant):
\begin{equation}\label{un2}
x=\frac{r}{r_0},\quad
E=\frac{\hbar^2k^2}{2mr^2_0},\quad
g=\frac{2mr^2_0U_0}{\hbar^2}.
\end{equation}

Substituting $\psi(r)=\chi(x)/x$, we arrive at
\begin{equation}\label{Sch2}
\frac{\rmd^2\chi(x)}{\rmd x^2}+k^2\chi(x)=g \chi(x) \cos_{\mu}{x},
\end{equation}
where we can interpret the right hand side as the inhomogeneous term
within the scattering problem at large~$x$.

To analyze the solution for an arbitrary deformation parameter $\mu>0$,
we turn to the general formalism, transforming the Schr\"odinger
equation into a matrix (integral) one.
We apply the orthogonal basis of spherical Bessel functions:
\begin{equation}
\int_0^\infty j_0(ux)\,j_0(vx)\,x^2\rmd x=\frac{\pi}{2uv}\delta(u-v),
\end{equation}
where $u$ and $v$ are real; the function $j_0(x)=\sin{x}/x$ such that
\begin{equation}
\Delta_x j_0(kx)+k^2j_0(kx)=0.
\end{equation}

Therefore, the following Hankel transform takes place:
\begin{eqnarray}
{\tilde f}(k)&=&\frac{2}{\pi}\int_0^\infty f(x)\,j_0(kx)\,x^2\rmd x,\label{HT}\\
f(x)&=&\int_0^\infty {\tilde f}(k)\,j_0(kx)\,k^2\rmd k.\label{iHT}
\end{eqnarray}

In dimensionless variables, Eq.~(\ref{Sch1}) is reduced to
the form:
\begin{eqnarray}\label{Sch3}
&&\hspace*{-1mm}(q^2-k^2){\tilde\psi}_k(q)+g\int_0^\infty U_\mu(q,q_*) {\tilde\psi}_k(q_*)\,q^2_*\rmd q_*=0,\\
&&\hspace*{-1mm}U_\mu(q,q_*)=\frac{2}{\pi}\int_0^\infty j_0(qx)\,\cos_\mu(x)\,j_0(q_*x)\,x^2\rmd x,
\end{eqnarray}
where $U(q,q_*)$ is the matrix element of the potential $\cos_\mu{x}$
and depends on absolute values of $q$ and $q_*$.

Neglecting the dependence on angle $\theta$ between wave vectors $\mathbf{q}$
and $\mathbf{q}_*$, $U(q,q_*)$ takes into account the first term of expansion:
\begin{eqnarray}
&&j_0(xQ(\theta))=\sum\limits_{\ell=0}^\infty(2\ell+1)j_\ell(qx)\,j_\ell(q_*x)\,P_\ell(\cos{\theta}),\\
&&Q(\theta)=\sqrt{q^2+q^2_*-2qq_*\cos{\theta}},
\end{eqnarray}
where $P_\ell(z)$ is the Legendre polynomial; $P_0(z)=1$.

In fact, the matrix $U(q,q_*)$ admits two fixed values of scattering angle $\theta$
which give us $Q(0)=|q-q_*|$ and $Q(\pi)=|q+q_*|$.

Our calculations yield:
\begin{eqnarray}
U_\mu(q,q_*)&=&\frac{1}{2qq_*}[T_\mu(q-q_*)-T_\mu(q+q_*)];\label{U1}\\
T_\mu(q)&\equiv&\frac{2}{\pi}\int_0^\infty \cos{(qx)}\,\cos_\mu{x}\,\rmd x\nonumber\\
&=&\frac{1}{\Gamma(1/\mu)}\left(\frac{|q|}{\mu}\right)^{\frac{1}{\mu}}\exp{\left(-\frac{|q|}{\mu}\right)}.
\label{Tq}
\end{eqnarray}
where it is convenient to substitute the expression in terms of
the Gauss hypergeometric function:
\begin{equation}
\cos_\mu{x}={}_2F_1\left(\frac{1}{2}+\frac{1}{2\mu},1+\frac{1}{2\mu};\frac{1}{2};-\mu^2x^2\right).
\end{equation}

The absolute value $|q|$ of a real $q$ in (\ref{Tq}) provides $T_\mu(-q)=T_\mu(q)$
in accordance with the definition. Note also that $U_\mu(q,-q_*)=U_\mu(q,q_*)$.

Besides, it is useful to define
\begin{equation}\label{qcos}
u_\mu(q)\equiv\lim\limits_{q_*\to0}U_\mu(q,q_*)
=\frac{q-1}{q^2\mu}\,T_\mu(q),
\end{equation}
in order to obtain an integral form of writing $\cos_\mu{x}$:
\begin{equation}
\cos_\mu{x}=\int_0^\infty u_\mu(q)\,j_0(qx)\,q^2\rmd q.
\end{equation}

To solve Eq.~(\ref{Sch3}), we substitute the expression~\cite{LL}:
\begin{equation}\label{ppsi}
{\tilde\psi}_k(q)=\frac{1}{q^2}\delta(q-k)+{\tilde\eta}_k(q),
\end{equation}
where, due to Eq.~(\ref{iHT}), the first term results in
$j_0(kx)$ that corresponds to the $s$-component of plane wave.

To soften the singularities of Eq.~(\ref{Sch3}), we set the unknown
function ${\tilde\eta}_k(q)$ as \cite{LL}
\begin{equation}\label{etaF}
{\tilde\eta}_k(q)=\frac{F(k,q)}{q^2-k^2}.
\end{equation}

The new function $F(k,q)$ must satisfy the equation:
\begin{equation}\label{Feq}
F(k,q)=-gU_\mu(k,q)-g\int_0^\infty\frac{F(k,q_*)U_\mu(q_*,q)}{q^2_*-k^2}\,q^2_*\rmd q_*.
\end{equation}

Restoring the dependence of wave function on space,
let us substitute (\ref{etaF}) into (\ref{iHT}) to obtain:
\begin{eqnarray}
\eta_k(x)&=&-\frac{\rmi}{2x}\int_0^\infty\frac{F(k,q)\rme^{\rmi qx}-F(k,-q)\rme^{-\rmi qx}}{q^2-k^2}
\,q\rmd q \nonumber\\
&=&-\frac{\rmi}{2x}\int_{-\infty}^\infty\frac{F(k,q)\rme^{\rmi qx}}{q^2-k^2}\,q\rmd q,
\label{eta2}
\end{eqnarray}
where we have used that $F(k,-q)=F(k,q)$.

\begin{figure}
\begin{center}
\includegraphics[width=5cm]{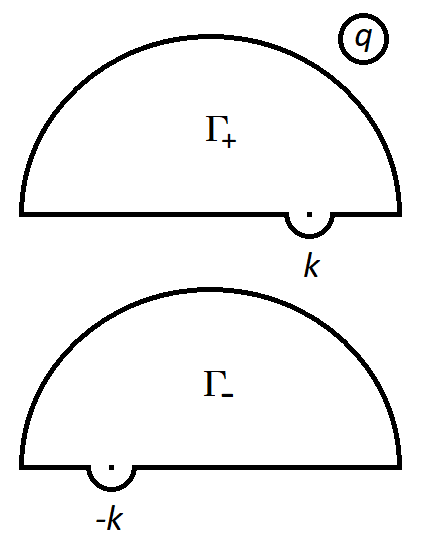}
\end{center}
\vspace*{-7mm}
\caption{Integration contours for calculating the wave function $\eta_k(x)$.
The contour $\Gamma_+$ takes into account a positive momentum $k$ of outgoing wave,
while $\Gamma_-$ is used for incoming wave.}\label{fig4}
\end{figure}

It is useful to decompose $\eta_k(x)$ into a sum of two integrals:
\begin{equation}\label{int2}
I_\pm(x)=-\frac{\rmi}{4x}\int_{-\infty}^\infty\frac{F(k,q)\rme^{\rmi qx}}{q\mp k}\,\rmd q.
\end{equation}

Computing the indefinite integrals (\ref{int2}), we treat $q$ as a complex variable
and choose the corresponding integration contours $\Gamma_\pm$ in the complex plane,
sketched in Fig.~\ref{fig4}, to provide decaying factor $\exp{(-x\,{\rm Im}\,q)}$.
The integrals are then determined by two residues (multiplied by $2\pi\rmi$), which
give us
\begin{equation}\label{eta3}
\eta_k(x)=\frac{\cos{(kx)}}{x}\,\frac{\pi}{2} F(k,k).
\end{equation}
Thus, this expression also consists of two terms which correspond to
the two scattering angles $\theta=0$ and $\theta=\pi$ in $q=k\cos{\theta}$.

As it follows from Eq.~(\ref{eta3}), the $s$-scattering amplitude is given by
expression $f(k)=(\pi/2) F(k,k)$, permitting us to find the scattering length $a$.

In an effort to obtain $F(k,k)$ as a series in powers of $g$, Eq.~(\ref{Feq})
can be solved by using an iterative procedure with the initial value
$F^{(0)}(k,q)=-gU_\mu(k,q)$. Limiting ourselves by this initial stage,
we obtain that
\begin{equation}
f^{(0)}(k)=g\frac{\pi\mu^{-\frac{1}{\mu}}}{4\Gamma(1/\mu)}(2|k|)^{\frac{1}{\mu}}k^{-2}\rme^{-\frac{2|k|}{\mu}}.
\end{equation}
Note the scaling law $f^{(0)}\sim k^{\frac{1}{\mu}-2}$ at small $k$.

Using that $k f^{(0)}(k)=\tan{\delta^{(0)}(k)}$, where $\delta^{(0)}(k)$
denotes the phase shift, we get $\psi_k(x)$ for large $x$ by summing
$j_0(kx)$ and $\eta_k(x)$ (see (\ref{ppsi})):
\begin{equation}\label{scatphi}
\psi_k(x)=C_k\frac{\sin{\left[kx+\delta^{(0)}(k)\right]}}{kx},
\end{equation}
where $C_k$ is a normalization.

The phase shift is evaluated as
\begin{equation}
\delta^{(0)}(k)=
\arctan{\left[\frac{g\pi}{4\Gamma(1/\mu)}\left(\frac{2}{\mu}\right)^{\frac{1}{\mu}}
k^{\frac{1}{\mu}-1}\rme^{-\frac{2k}{\mu}}\right]}.
\end{equation}
We note a weak dependence of this on small $k$ at $\mu=1$:
\begin{equation}\label{d1}
\delta^{(0)}_{\mu=1}(k)=\arctan{\left(\frac{g\pi}{2}\rme^{-2k}\right)},
\end{equation}
when $V_{\mu=1}(r_0x)\simeq(-g)\, x^{-2}$ at $x\gg1$, see Fig.~\ref{fig3}(b).

Let us relate the real-valued amplitude $f^{(0)}(k)$ with the scattering
amplitude $f_{\rm B}(2k\sin{(\theta/2)})$ in the Born approximation,
where $f_{\rm B}(q)=-(g\pi/2)\,u_\mu(q)$ accordingly to its
definition~\cite{LL} and Eq.~(\ref{qcos}). In fact,
we need to extract $f^{(0)}(k)\sim P_0(\cos{\theta})$ from $f_{\rm B}$,
using the orthogonality property of the Legendre polynomials $P_\ell(z)$.

One has
\begin{equation}
f^{(0)}(k)=\frac{1}{2}\int_0^\pi f_{\rm B}\left(2k\sin{\frac{\theta}{2}}\right)
\,\sin{\theta} \rmd\theta,
\end{equation}
where we have substituted $P_0(\cos{\theta})=1$. This formula is also confirmed
numerically.

Thus, the amplitude $f^{(0)}(k)$ is the $s$-component of the Born
amplitude, and then we can evaluate the scattering length (in units of $r_0$)
as a function of $\mu$ using the relation:
\begin{equation}\label{sla}
a^{(0)}=-\lim_{k\to0}f^{(0)}(k).
\end{equation}

We have that $a^{(0)}\to0$ for $\mu<1/2$; $a^{(0)}=-4g\pi$ at $\mu=1/2$;
$a^{(0)}\to-\infty$ for $\mu>1/2$. Therefore, the potential $\cos_\mu(r/r_0)$
demonstrates a long-range attraction for $\mu\geq1/2$ and reveals the unitary
regime with $1/a^{(0)}=0$ if $\mu>1/2$.

\subsection{Case of $\mu=1$}

\begin{figure}
\begin{center}
\includegraphics[width=7.8cm]{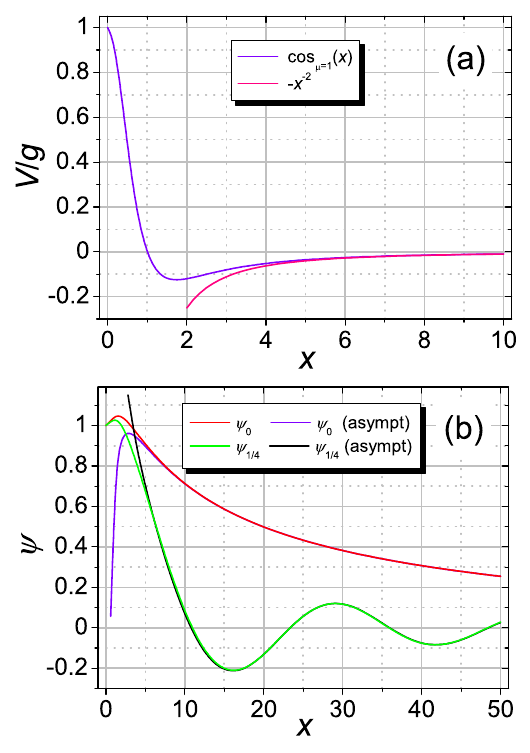}
\end{center}
\vspace*{-7mm}
\caption{Panel (a): Interparticle potential $\cos_{\mu=1}{x}$ (violet line) and
$-1/x^2$ (pink line). Panel (b): Wave functions $\psi_k$ and their asymptotics at
$\mu=1$ for different $k$: red and violet lines for $k=0$; green and black lines
for $k=1/4$. The fixed parameters are $g=1/2$ and $c_{1/2}=1$.}\label{fig5}
\end{figure}

Let us look at the case of $\mu=1$ in more detail, when
\begin{equation}\label{cos1}
\cos_{\mu=1}{x}=\frac{1-x^2}{(1+x^2)^2}.
\end{equation}
This function varies within the range $[1;-1/8]$, see Fig.~\ref{fig5}(a).

We turn again to the Schr\"odinger equation (\ref{Sch2}), imposing conditions
$\chi(0)=0$ and $\chi^\prime(0)=c_g$, in order to get $\psi=\chi(x)/x$.
Then the analytical unnormalized solution for $\mu=1$ is given by
\begin{eqnarray}
\chi_k(x)&=&c_g\,x\,(1+x^2)^{\frac{1}{2}+\frac{1}{2}\beta}\times\nonumber\\
&&\times H_C\left(0,\frac{1}{2},\beta,-\frac{k^2}{4},\frac{k^2}{4}-\frac{g}{4}+\frac{1}{2},-x^2\right),
\label{chi1}
\end{eqnarray}
here $\beta=\sqrt{1-2g}$.

The Heun confluent function $H_C(\alpha,\beta,\gamma,\delta,\eta,z)$
(altogether with $z^{-\beta}H_C(\alpha,-\beta,\gamma,\delta,\eta,z)$)
is a local Frobenius solution of the problem:
\begin{eqnarray}
&&y^{\prime\prime}+\left(\alpha+\frac{\beta+1}{z}+\frac{\gamma+1}{z-1}\right)y^\prime
+\left(\frac{\nu}{z-1}+\frac{\mu}{z}\right)y=0,\nonumber\\
&&\delta=\mu+\nu-\alpha\frac{\beta+\gamma+2}{2},
\nonumber\\
&&\eta=\alpha\frac{\beta+1}{2}-\frac{\beta+\gamma+\beta\gamma}{2}-\mu.
\nonumber
\end{eqnarray}

Particularly, we would like to note special unnormalized solution $\phi_{+}(x)$ at $g=3/8$:
\begin{eqnarray}
&&-\frac{\rmd^2\phi_{+}}{\rmd x^2}+\frac{3}{8}\frac{1-x^2}{(1+x^2)^2}\phi_{+}=0,\quad
\phi_{+}(0)=0,\label{phi+}\\
&&\phi_{+}(x)=A_{+}(1+x^2)^{1/4}\sin{\left[\frac{\sqrt{2}}{4}\ln{\left(x+\sqrt{1+x^2}\right)}\right]}.
\nonumber
\end{eqnarray}
The total wave function is then $\psi_{+}=\phi_{+}(x)/x$.


For small positive $k$ (and $g$), we construct oscillating asymptotic solution
using phase shift (\ref{d1}):
\begin{equation}\label{as1}
\psi^{\rm as}_k(x)=\frac{A_k}{x}\sin{\left[kx+\arctan{\left(\frac{g\pi}{2}\rme^{-2k}\right)}\right]},
\end{equation}
where $A_k$ is found from the equality $\psi^{\rm as}_k(x_0)=\chi_k(x_0)/x_0$
for some fixed $x_0\gg1$.

The case $k=0$ requires separate consideration. To describe the asymptotic
in this situation, we appeal to the auxiliary equation:
\begin{equation}
-\frac{\rmd^2\chi^{\rm as}_0}{\rmd x^2}-\frac{g}{x^2}\chi^{\rm as}_0=0.
\end{equation}
Its solutions are $x^{\alpha_+}$ and $x^{\alpha_-}$
with $\alpha_\pm=(1\pm\sqrt{1-4g})/2$.
Then a stable and finite solution can be guaranteed at $g\geq1/4$. Substituting
$g=1/4+s^2$, $s\in\mathbb{R}_+$, and combining linearly $x^{\alpha_\pm}$,
we arrive at $\psi^{\rm as}_0=\chi^{\rm as}_0/x$:
\begin{equation}\label{as2}
\psi^{\rm as}_0(x)=\frac{A_0}{\sqrt{x}}\sin{(s\ln{x}+d_0)},
\end{equation}
where $A_0$ and $d_0$ are arbitrary constants.

It is useful to compare (\ref{phi+}) at $g=3/8$ with $\chi^{\rm as}_0$.
Indeed, function $\phi_{+}(x)$ for $x\gg1$ tends to $\chi^{\rm as}_0(x)$ at
$s=\sqrt{2}/4$ and $d_0=s\ln{2}$.

We would like to test the asymptotic (\ref{as2}) by determining the parameters
$A_0$ and $d_0$ on the base of two fixed points of the exact curve $\psi_0(x)$.
It leads to the equations:
\begin{eqnarray}
&&\hspace*{-2mm}
\arcsin{\frac{\sqrt{x_2}\,\psi_0(x_2)}{A_0}}
-\arcsin{\frac{\sqrt{x_1}\,\psi_0(x_1)}{A_0}}=s\ln{\frac{x_2}{x_1}},\label{app1}\\
&&\hspace*{-2mm}
d_0=\arcsin{\frac{\sqrt{x_2}\,\psi_0(x_2)}{A_0}}-s\ln{x_2}.\label{app2}
\end{eqnarray}
The first transcendent equation determines $A_0$, while the latter
returns the value of phase $d_0$. Note that $d_0$ and $A_0$ are not
independently fixed.

Then, the admissible behavior of two particles is described by the wave functions
as in Fig.~\ref{fig5}(b), where $g=1/2$ and $s=1/2$ for definiteness. It is seen
that their maximum values are located near the potential minimum (see Fig.~\ref{fig5}(a))
and depends on the energy (or wave number $k$). A higher energy
value allows particles to approach each other, overcoming repulsion at short
distances.

We see in Fig.~\ref{fig5}(b) that the asymptotics (\ref{as1}) and (\ref{as2}) are
valid at $x\gg1$. In Eq.~(\ref{as1}) we have used the phase shift (\ref{d1})
which is derived analytically. The amplitude $A_{1/4}=3.538$ for $k=1/4$ is
determined at $x_0=20$.
To fix the parameters of (\ref{as2}), we choose the
points $x_1=20$, $x_2=30$ and obtain $A_0=-2.275$ and $d_0=-2.867$.

We provide the applicability of the phase shift $\delta^{(0)}(k)$ in
the Born approximation by the smallness of the coupling $g<1$. However,
dealing with a problem with an infinite scattering length $a$,
the dependence of phase shift on running $g$ can be more complicated~\cite{BrH06}.

Besides, the accurate derivation of the phase shift $d_0$ still requires
a detailed study. We only note the connection between the phase
of the wave function (\ref{as2}) and the phase in semiclassical
approximation (with a non-zero orbital number $l$ in general) \cite{LL}:
\begin{equation}
\int_{x_0}^x\sqrt{k^2-\frac{(l+1/2)^2}{x^2}-g\cos_{\mu}{x}}\,\rmd x+\frac{\pi}{4},
\end{equation}
where $x_0$ is the zero of the expression under square root.

Setting $l=0$, $k=0$, and $\cos_{\mu=1}{x}\simeq-x^{-2}$
for large $x$, this phase takes the form
\begin{eqnarray}
&&s\ln{x}-s\ln{x_0}+\frac{\pi}{4},\\
&&x_0=\sqrt{\frac{2g+1+2\sqrt{g^2+2g}}{4g-1}},\quad
s=\sqrt{g-1/4}.\nonumber
\end{eqnarray}
Although it may need some improvement to apply.

Let us consider a problem with a negative coupling constant $g=-|g|$,
when there are an attractive well in the range $x\in[0;1]$ and
a positive-valued barrier (with a maximum $|g|/8$) for $x>1$.
At zero energy ($k=0$), the barrier width and positive scattering
length $a$ become infinite.

\begin{figure}
\begin{center}
\includegraphics[width=7cm]{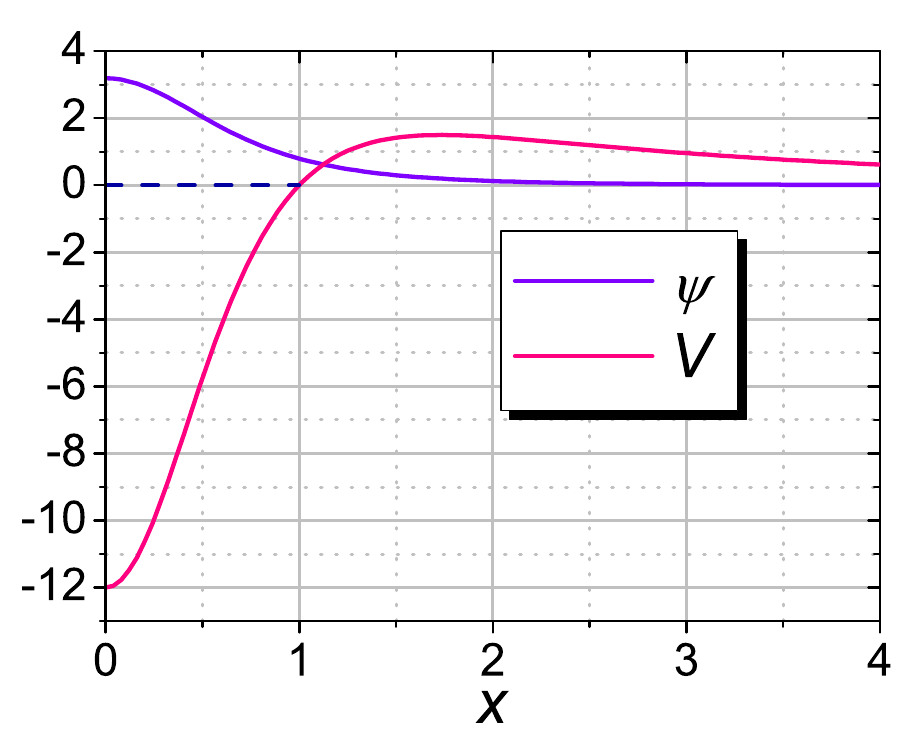}
\end{center}
\vspace*{-7mm}
\caption{Interparticle potential $V=-12\cos_{\mu=1}{x}$ (pink line) and
normalized wave function (violet line) as functions of interparticle distance $x$.
Dashed line for $x\in[0;1]$ indicates classically admissible region of the problem
(\ref{fm1}).}\label{fig6}
\end{figure}

Analyzing the solution (\ref{chi1}), we reveal a stable, physically
correct solution $\phi_{-}(x)$ in the case of $|g|=12$. This unique
value of $g$ simplifies the problem considerably:
\begin{eqnarray}
&&-\frac{\rmd^2\phi_{-}}{\rmd x^2}-12\,\frac{1-x^2}{(1+x^2)^2}\,\phi_{-}=0,\quad
\phi_{-}(0)=0,\label{fm1}\\
&&\phi_{-}(x)=\sqrt{\frac{2}{\pi}}\frac{4x}{(1+x^2)^2},\quad
\int_0^\infty |\phi_{-}(x)|^2\rmd x=1.\label{fm2}
\end{eqnarray}
Function $\phi_{-}(x)$ determines $\psi_{-}=\phi_{-}(x)/x$.

The dependence presented in Fig.~\ref{fig6} allows us to interpret the obtained
solution as a two-particle ``molecule'' in the ground state without engaging
any chemical bond in the usual sense. We also have $\langle x^2\rangle=1$ that
corresponds to classically admissible region.

We get also unnormalized solution for non-zero energy $E=k^2\geq0$ in the terms
of Heun function:
\begin{equation}
\chi_k(x)=\frac{x}{(1+x^2)^2}\,H_C\left(0,\frac{1}{2},-5,-\frac{k^2}{4},\frac{7}{2}+\frac{k^2}{4},-x^2\right),
\end{equation}
which generalizes $\phi_{-}(x)$ and satisfies the equation:
\begin{equation}
-\frac{\rmd^2\chi_k}{\rmd x^2}-12\,\frac{1-x^2}{(1+x^2)^2}\,\chi_k=k^2 \chi_k,\quad
\chi_k(0)=0.
\end{equation}

In the energy range $0<E<E_{\rm thr}$, where the threshold $E_{\rm thr}=3/2$
corresponds to the potential maximum, the wave function $\psi_k(x)=\chi_k(x)/x$
has a long-wavelength tail with a large but finite amplitude, and
its shape in the interval $x\in[0;1]$ does not change radically in
comparison with Fig.~\ref{fig6}. In fact, it describes an excited system.

On the other hand, the function $\psi_k(x)$ at $E>E_{\rm thr}$ has
an oscillating damped tail which is typical for scattering.

Thus, we identify the problem (\ref{fm1})--(\ref{fm2}) with the description
of DM dimer. We admit that a similar solution exists for $\mu<1$.
However, in the case of $\mu=1$, it has the simplest form suitable
for further use.

\section{\label{S4}Dimer Creation Due to Resonance Scattering}

Actually, resonance scattering implies the detection of (quasi)discrete
energy levels admitted by the interaction potential. Indeed, it would be
interesting to take into account (quasi)bound states in scattering for
a number of deformed potentials $\cos_{\mu}(x)$, but we do not touch on this.
Here the (quasi)discrete level arise differently.

This study is addressed to the last problem of the previous section, which
suggests the existence of a ``molecule'' of two DM particles. We would like
to consider in detail the mechanism of its formation, because one good
potential is not enough to form a bound state of DM particles.

Our approach inherits ideas of the Feshbach resonance~\cite{Fesh62}, which uses
at least two channels of scattering, one open and one closed channels. In fact,
it allows one to create (quasi)discrete level and turn a pair of scattered particles
into a ``molecule'' and vice versa under auxiliary influence. Despite questions
about the nature of interactions in DM, such an approach seems to be appropriate,
since it looks difficult to bring directly a particle with zero energy into
the scope of potential $-12\cos_{\mu=1}(x)$ of another particle.

We assume that a pair of spinless particles in a collision jumps between open and
closed channels.
When the total energy exceeds the open channel threshold ($E=0$), the open channel in
such a system is both an incoming and an outgoing channel.
Scattering reveals the Feshbach resonance when the energy of the bound state of
the closed channel is close to the threshold of the open channel. Due to the coupling
of the channels, the unperturbed bound state of the closed channel becomes dressed,
taking into account an interaction. This dressed state is treated as
a (quasi)bound state of the entire scattering system. Scattered particles temporarily
pass into a (quasi)bound state and return to the open channel after a characteristic
time $\tau=2/\Gamma$, determined by the decay width $\Gamma$ of the (quasi)bound state.

Formulating our model based on the stationary Schrödinger equation for $s$-wave
functions, we fix one particle at the point $x=0$, and the partner particle is moving
and is characterized by the relative radial coordinate $x\geq0$. We assume that this
particle is in one of two channels. A closed channel in the absence of external
interaction describes a molecule in the ground state and has energy $E^{(1)}=0$,
and an open channel corresponds to elastic wave scattering due to another interaction
and initially has a small energy $E^{(2)}=k^2$ determined by the relative momentum $k$.
Between these channels there is an energy gap $Q=E^{(2)}-E^{(1)}>0$, which can be reduced
with the help of external influence. In general, there are three different interactions
in such a system, and $E^{(1)}$ and $E^{(2)}$ are not energy levels of the same potential.
For this reason, the interaction parameters must be tuned to obtain the desired effect.

We describe the relative motions of a particle in two channels, adopting
the Fock-space representation~\cite{Yam93}:
\begin{equation}\label{Meq1}
\mathbb{H} X=E X,\qquad
\mathbb{H}=
\left(
\begin{array}{cc}
H_{\rm m} & W \\
W^\dag & H_{\rm wv}
\end{array}
\right),
\end{equation}
where $H_{\rm m}$ and $H_{\rm wv}$ are the Hamiltonians of
the molecule (closed channel) and the scattered wave (open channel),
respectively. The coupling between channels is represented by $W$,
and it is associated with extra force.

We account for the energy gap $Q$ in Eq.~(\ref{Meq1}) by defining
the Hamiltonian of molecule as $H_{\rm m}=H_1+Q$, where
\begin{equation}
H_1=-\frac{\rmd^2}{\rmd x^2}-12\,\frac{1-x^2}{(1+x^2)^2}.
\end{equation}

In a sense, the system is doubly degenerate at $W=0$ due to existing two
independent wave functions for the same eigenvalue $E$:
\begin{equation}\label{P0}
X^{(1)}=\left(
\begin{array}{c}
\chi^{(1)}\\
0
\end{array}
\right),\qquad
X^{(2)}=
\left(
\begin{array}{c}
0\\
\chi^{(2)}
\end{array}
\right),
\end{equation}
which are evidently orthogonal in this representation.
We can immediately identify $\chi^{(1)}$ with $\phi_{-}$ from
Eq.~(\ref{fm2}). In principle, we should write
$X=X^{(1)} \cos{\alpha}+X^{(2)} \sin{\alpha}$ with some $\alpha$
to normalize the total wave function $X$ with the respect to
Fock representation.

As shown in the previous section, the essential spatial interval
for the existence of a molecule is $x\in[0;1]$. Therefore, significant
processes take place in this region, which we call the resonance zone.
For this reason, we concentrate the external force, parametrized
by $\omega$, there:
\begin{equation}\label{exi}
W(x)=-\omega^2\,\theta(1-x),\qquad W^\dag=W.
\end{equation}

Similarly, we describe an interaction in the open channel by using
the square well potential for simplicity:
\begin{equation}
V_{\rm sq}(x)=-V\,\theta(1-x),\quad
V>0.
\end{equation}

In fact, all quantities in such a system are divided into two components
belonging either to interval $x\in[0;1]$ or to interval $x\in[1;\infty)$,
which we label as ``$<$'' and ``$>$'' with the respect to separating point
$x=1$, respectively. Then, the wave functions in the channels numbered by
$\alpha=1,2$ are decomposed as
\begin{equation}
\chi^{(\alpha)}(x)=\theta(1-x)\,\chi^{(\alpha)}_{<}(x)+\theta(x-1)\,\chi^{(\alpha)}_{>}(x).
\end{equation}

We require that $\chi^{(\alpha)}_{<}(0)=0$, and connect functions
in separating point $x=1$ by the merging condition:
\begin{equation}\label{rBC}
\left.\frac{\rmd}{\rmd x}\,\ln{\chi^{(\alpha)}_{<}(x)}\right|_{x=1}=
\left.\frac{\rmd}{\rmd x}\,\ln{\chi^{(\alpha)}_{>}(x)}\right|_{x=1},
\end{equation}
to guarantee an equality of derivatives and proportionality of
the functions in the left and right sides of (\ref{rBC}).

Before proceeding, we recall the known results for the open channel
($\alpha=2$) in the absence of coupling $W$. The scattering characteristics
follow from the equation:
\begin{equation}\label{sq1}
H_{\rm wv} \chi^{(2)}\equiv\left(-\frac{\rmd^2}{\rmd x^2}+V_{\rm sq}(x)\right) \chi^{(2)}
=E \chi^{(2)}.
\end{equation}

As mentioned above, solution of Eq.~(\ref{sq1}) is given by
\begin{eqnarray}
&&\hspace*{-2mm}
\chi^{(2)}(x)=\theta(1-x)\,\chi^{(2)}_{<}(x)+\theta(x-1)\,\chi^{(2)}_{>}(x),\label{ch2}\\
&&\hspace*{-2mm}
\chi^{(2)}_{<}(x)=\frac{\sin{Kx}}{K},\quad\
\chi^{(2)}_{>}(x)=\frac{\sin{K}}{K}\,\frac{\sin{(kx+\delta)}}{\sin{(k+\delta)}},
\label{unp2}
\end{eqnarray}
where $K=\sqrt{E+V}$ and $k=\sqrt{E}$. When $E<0$, $\chi^{(2)}_{>}$ behaves
as $\exp{(-x\sqrt{|E|})}$.
The phase shift $\delta$ is derived from the merging condition (\ref{rBC}):
\begin{equation}\label{screl1}
K \cot{K}=k \cot{(\delta+k)}.
\end{equation}

The scattering length~$a_V$ of potential~$V_{\rm sq}(x)$ is
\begin{equation}\label{aV}
a_V=1-\frac{\tan{\sqrt{V}}}{\sqrt{V}}.
\end{equation}
Note that $a_V$ demonstrates discontinuous behavior at
$\sqrt{V}=\pi/2$. We exclude from our consideration this
``zero energy resonance''.

The scattering matrix $S=\rme^{2\rmi\delta}$ and amplitude $f$ are
\begin{equation}\label{scam1}
S=\rme^{-2\rmi k}\,\frac{K \cot{K}+\rmi k}{K \cot{K}-\rmi k},\qquad
f=\frac{\rme^{2\rmi\delta}-1}{2\rmi k}.
\end{equation}

Thus, we admit a single bound state in the closed channel and a continuum
of spherical waves of relative momentum $k$ between the two particles in
the open channel. Taking into account the complexity of the molecule potential
for calculations, we describe the Feshbach resonance in the first approximation.

The set of equations in whole admissible space is
\begin{eqnarray}
&&(H_1+Q-E)\chi^{(1)}+W\chi^{(2)}=0,\label{e1-0}\\
&&(H_{\rm wv}-E)\chi^{(2)}+W^\dag \chi^{(1)}=0.\label{e2-0}
\end{eqnarray}

For convenience, we further use the bra- and ket-vectors to simplify
the notation of matrix elements.

In the first approximation, we put~\cite{Joa75}:
\begin{equation}\label{anz1}
|\chi^{(1)}\rangle=\lambda|\phi_{-}\rangle,
\end{equation}
where $\lambda$ is a complex constant which should be found.

Taking into account (\ref{anz1}) and acting by $\langle\phi_{-}|$
on Eq.~(\ref{e1-0}), one has
\begin{equation}
\lambda=\frac{\langle\phi_{-}|W|\chi^{(2)}_{<}\rangle}{E-Q},
\end{equation}
where we have used the equation $H_1|\phi_{-}\rangle=0$, the normalization
$\langle\phi_{-}|\phi_{-}\rangle=1$ and the equality
$\langle\phi_{-}|W|\chi^{(2)}\rangle=\langle\phi_{-}|W|\chi^{(2)}_{<}\rangle$
due to the form of $W(x)$. At this stage, the coefficient $\lambda$ depends on
the unknown function $\chi^{(2)}_{<}$.

Introducing the Hamiltonian:
\begin{equation}
H_2=-\frac{\rmd^2}{\rmd x^2}-K^2,\qquad
K^2=E+V,
\end{equation}
the equations for the open channel takes on the form:
\begin{eqnarray}
&&H_2|\chi^{(2)}_{<}\rangle+W^\dag|\chi^{(1)}\rangle=0,\label{e2-p1}\\
&&-\frac{\rmd^2\chi^{(2)}_{>}}{\rmd x^2}=k^2\chi^{(2)}_{>}.\label{e2-p2}
\end{eqnarray}

Solution to Eq.~(\ref{e2-p1}) can be written as
\begin{eqnarray}
|\chi^{(2)}_{<}\rangle&=&|\tau_0\rangle-G^{(+)}_2W^\dag|\chi^{(1)}\rangle
\nonumber\\
&=&|\tau_0\rangle-\lambda G^{(+)}_2W^\dag|\phi_{-}\rangle,
\label{e2-3}
\end{eqnarray}
where the unperturbed wave function $\tau_0(x)=\langle x|\tau_0\rangle$
is equal to $\chi^{(2)}_{<}(x)$ from Eq.~(\ref{unp2}) and is found as
\begin{equation}
H_2\,\tau_0(x)=0,\qquad
\tau_0(x)=\frac{\sin{Kx}}{K}.
\end{equation}

The Hamiltonian $H_2$ determines also the Green's operator
$G^{(+)}_2=(H_2-\rmi\epsilon)^{-1}$ that contains the shifted
energy $E+\rmi\epsilon$ at $\epsilon\to0$.
The corresponding Green's function is
\begin{eqnarray}
&&\hspace*{-4.2mm}
G^{(+)}_2(x,y;K)=\frac{\sin{Kx_{1}} \cos{Kx_{2}}}{K}
+\rmi\, \frac{\sin{Kx} \sin{Ky}}{K},\\
&&\hspace*{-4mm}
x_{1}={\rm min}(x,y),\qquad x_{2}={\rm max}(x,y),
\nonumber
\end{eqnarray}
which serves for finding the outgoing wave under the boundary
condition $G^{(+)}_2(0,y;K)=0$ to obtain a finite radial
solution for the $s$-wave. Accounting for other partial waves
in the Green's function may require the use of expression
$\exp{(\rmi K|{\bf x}-{\bf y}|)}/K$ resulted from the Helmholtz equation.

To express $\lambda$ in the terms of known solutions $\tau_0$
and $\phi_{-}$, let us act by $\langle\phi_{-}|W$ on Eq.~(\ref{e2-3}).
Then, it is easy to obtain that
\begin{equation}
\lambda=\frac{\langle\phi_{-}|W|\tau_0\rangle}
{E-Q+\langle\phi_{-}|WG^{(+)}_2W^\dag|\phi_{-}\rangle}.
\end{equation}
The condition of zeroing the denominator indicates the resonant state
of the compound system, i.e., the energy of a dressed (quasi)discrete
level~\cite{Fesh62,Joa75}.

It is useful to introduce the notations:
\begin{eqnarray}
\omega^4\Delta(K)&=&{\rm Re}\,\langle\phi_{-}|WG^{(+)}_2W^\dag|\phi_{-}\rangle,
\label{Del}\\
\omega^4\gamma(K)&=&{\rm Im}\,\langle\phi_{-}|WG^{(+)}_2W^\dag|\phi_{-}\rangle.
\label{gam}
\end{eqnarray}

To sketch how this works for a fixed $Q>0$, let us imagine
the situation when $V\gg E$ at $E\to0$, and the denominator of
$\lambda$ vanishes at some complex value of the energy
$E_0-\rmi\Gamma_0/2$, thereby making the magnitude of the wave functions
extremely large. The positive resonance energy $E=E_0$ and
decay width $\Gamma_0$ may be simply determined from the relations:
\begin{equation}\label{Rch}
E_0=Q-\omega^4\Delta(\sqrt{V}),\qquad
\Gamma_0=2\omega^4\gamma(\sqrt{V}).
\end{equation}
This shows that $Q>E_0>0$ due to two
additional interactions, which are controlled by $V$ and $\omega^2$.
Besides, the lifetime of a dimer in such a scenario would be
$\tau=2/\Gamma_0$ (in dimensionless units).

In general, we write that
\begin{equation}
\lambda=\frac{\langle\phi_{-}|W|\tau_0\rangle}
{E-Q+\omega^4\Delta(K)+\rmi \omega^4\gamma(K)}.
\end{equation}

Let us pay attention to finding exact expressions, starting with the overlap
integral that defines $\langle\phi_{-}|W|\tau_0\rangle$:
\begin{eqnarray}
B(K)&\equiv&\int_0^1 \frac{4x}{(1+x^2)^2}\,\frac{\sin{Kx}}{K}\,\rmd x\\
&=&[\pi+\rmi\, {\rm Ci}(K+\rmi K)-\rmi\, {\rm Ci}(K-\rmi K)] \cosh{K}
\nonumber\\
&-&[{\rm Si}(K+\rmi K)+{\rm Si}(K-\rmi K)] \sinh{K}-\frac{\sin{K}}{K}.
\nonumber
\end{eqnarray}

Here we have used the sine and cosine integrals:
\begin{equation}
{\rm Si}(z)=\int_0^z\frac{\sin{t}}{t}\rmd t,\ \
{\rm Ci}(z)=\gamma+\ln{z}+\int_0^z\frac{\cos{t}-1}{t}\rmd t,
\nonumber
\end{equation}
$\gamma=0.57721...$ is the Euler's constant.

We obtain that
\begin{equation}
\langle\phi_{-}|W|\tau_0\rangle=-\sqrt{\frac{2}{\pi}}\,\omega^2 B(K).
\end{equation}

Note that $B(K)$ is a decreasing oscillating function that
vanishes at $K_1=5.359982109$, $K_2=7.917546061$,
$K_3=11.09524951$ and so on. At these points one gets $\lambda=0$
and $\chi^{(2)}_{<}(x)=\tau_0(x)$ with $\kappa_n=-K_n \cot{K_n}>0$
on the boundary. Index $n$ corresponds to the number of zeros of
the wave function $\tau_0(x)$ in interval $x\in[0;1]$.
This indicates the possibility of appearing
virtual bound states in the system. However, we ignore them here,
focusing on the Feshbach resonance.

\begin{figure}
\begin{center}
\includegraphics[width=7.5cm]{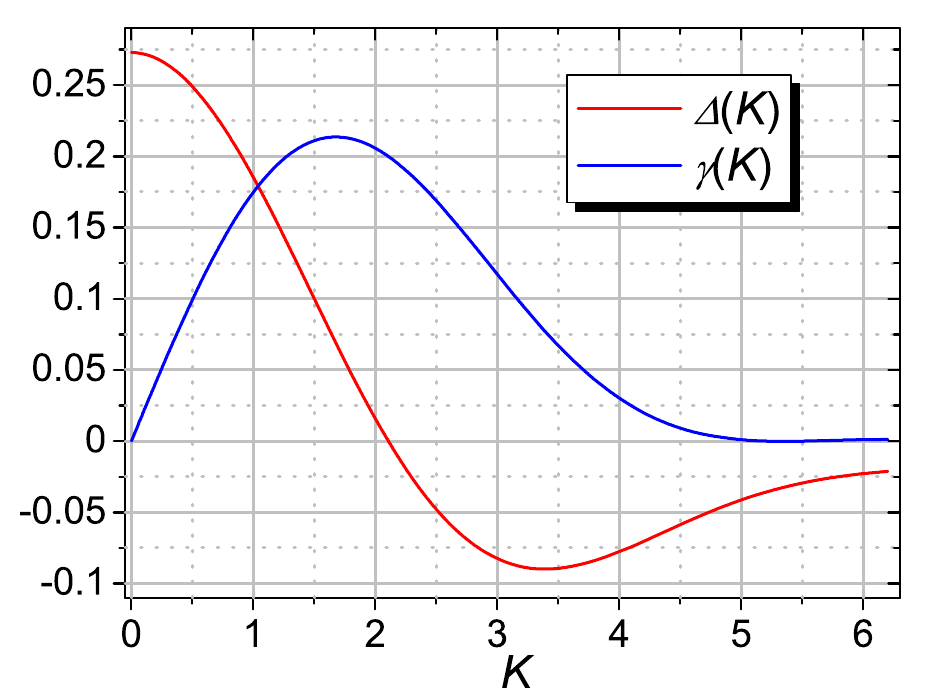}
\end{center}
\vspace*{-7mm}
\caption{The functions $\Delta(K)$ and $\gamma(K)$ which determine
$\langle\phi_{-}|WG^{(+)}_2W^\dag|\phi_{-}\rangle$. In the given
interval there is zero $\Delta(K_0)=0$ for $K_0=2.107331071$,
while $\gamma(K_0)=0.200330$.}\label{fig7} 
\end{figure}

The first correction to the wave function $\chi^{(2)}_{<}(x)$
is determined by $G^{(+)}_2W^\dag|\phi_{-}\rangle$.
To find the complex function $\langle x|G^{(+)}_2W^\dag|\phi_{-}\rangle=-\omega^2\tau_1(x)$,
we turn to the solution of the inhomogeneous equation:
\begin{eqnarray}
&&H_2\,\tau_1(x)=\phi_{-}(x),\\
&&\tau_1(x)=\int_0^1 G^{(+)}_2(x,y;K)\,\phi_{-}(y)\,\rmd y.
\end{eqnarray}

Computing, the components of $\tau_1(x)=\tau^{\rm R}_1(x)+\rmi\,\tau^{\rm I}_1(x)$
are written as
\begin{eqnarray}
\sqrt{\frac{\pi}{2}}\,\tau^{\rm R}_1(x)&=&\rmi[\zeta(Kx+\rmi K)-\zeta(Kx-\rmi K)]
\nonumber\\
&&-\left[\varsigma_{+}(K)+\varsigma_{-}(K)+\frac{\cos{K}}{K}\right] \sin{Kx}
\nonumber\\
&&+\pi \cosh{K} \cos{Kx},\\
\sqrt{\frac{\pi}{2}}\,\tau^{\rm I}_1(x)&=&B(K) \sin{Kx},
\end{eqnarray}
where we denote
\begin{eqnarray}
\zeta(z)&=&{\rm Ci}(z) \cos{z}+{\rm Si}(z) \sin{z},
\nonumber\\
\varsigma_{\pm}(K)&=&{\rm Ci}(K\pm\rmi K) \sinh{K}\pm\rmi\, {\rm Si}(K\pm\rmi K) \cosh{K}.
\nonumber
\end{eqnarray}

Taking into account the form of $\tau^{\rm I}_1(x)$ and the definition of $B(K)$,
we specify the function $\gamma(K)$ (see Eq.~(\ref{gam})):
\begin{eqnarray}
\gamma(K)&=&\frac{2}{\pi}\,K B^2(K);\\
\gamma(K)&\simeq&\frac{(\pi-2)^2}{2\pi}\,K,\quad K\to0.
\end{eqnarray}

At this time, the real part $\tau^{\rm R}_1(x)$ determines also a deviation
$\Delta(K)$ of the resonance energy:
\begin{equation}
\Delta(K)=\int_0^1 \phi_{-}(x)\,\tau^{\rm R}_1(x)\,\rmd x.
\end{equation}
This integral is not simple to be calculated analytically, and we present
the numerical result in Fig.~\ref{fig7}.

Combining, the wave function in the first approximation for the closed channel is
\begin{equation}\label{ss1}
\chi^{(1)}(x)=-\frac{\omega^2}{D(K)}\,\sqrt{\frac{\gamma(K)}{K}}\,\phi_{-}(x),
\end{equation}
where we have introduced the notation:
\begin{eqnarray}
D(K)&=&K^2-V-Q+\omega^4\Delta(K)+\rmi \omega^4\gamma(K)
\nonumber\\
&=&D_{\rm R}(K)+\rmi D_{\rm I}(K).\label{deno}
\end{eqnarray}

Solution~(\ref{ss1}) describes the short-living state of dimer
and vanishes at $\omega=0$. Beyond the resonance zone we arrive at equation
$(H_1+Q-E)\chi^{(1)}_{>}=0$ for $Q=E$ as it was assumed.

Due to proportionality $\tau^{\rm I}_1(x)\sim\tau_0(x)$, we present
the open channel wave function in the resonance zone as
\begin{equation}
\chi^{(2)}_{<}(x)=\frac{D_{\rm R}(K)}{D(K)}\,\tau_0(x)
-\frac{\omega^4}{D(K)}\,\sqrt{\frac{\gamma(K)}{K}}\,\tau^{{\rm R}}_{1}(x).
\label{ss2<}
\end{equation}

To find the phase shift $\delta$ for the wave function $\chi^{(2)}_{>}(x)$:
\begin{eqnarray}
\chi^{(2)}_{>}(x)&=&\chi^{(2)}_{<}(1)\,\frac{\sin{(kx+\delta)}}{\sin{(k+\delta)}},\\
\chi^{(2)}_{<}(1)&=&\frac{D_{\rm R}(K) \sin{K}-D_{\rm I}(K) \cos{K}}{K\,D(K)},
\end{eqnarray}
we appeal to the merging condition (\ref{rBC}) which gives us
\begin{equation}\label{BC2}
K\,\frac{D_{\rm R}(K) \cos{K}+D_{\rm I}(K) \sin{K}}{D_{\rm R}(K) \sin{K}-D_{\rm I}(K) \cos{K}}
=k \cot{(k+\delta)}.
\end{equation}

This relation can be re-written as
\begin{eqnarray}
&&K \cot{(K-\delta_{\rm rs})}=k \cot{(k+\delta)},\\
&&\delta_{\rm rs}(K)=\arctan{\frac{D_{\rm I}(K)}{D_{\rm R}(K)}},
\end{eqnarray}
where $\delta_{\rm rs}$ is the phase shift which is caused by
complex interaction in the resonance zone. At $\delta_{\rm rs}\equiv0$,
only potential scattering with $V_{\rm sq}(x)$ remains.

It is easy to derive that
\begin{eqnarray}
\delta&=&-k+\arctan{\left[\frac{k}{K}\,\tan{(K-\delta_{\rm rs})}\right]},\\
&\simeq&k\left[\frac{\tan{(K-\delta_{\rm rs})}}{K}-1\right],
\end{eqnarray}
where the last expression is valid for small $k$.

The scattering matrix $S=\rme^{2\rmi\delta}$ for the open channel is
\begin{equation}
S=\rme^{-2\rmi k}\,\frac{K \cot{(K-\delta_{\rm rs})}+\rmi k}
{K \cot{(K-\delta_{\rm rs})}-\rmi k},\quad
K=\sqrt{k^2+V}.
\end{equation}
Its pole, resulted from zero of denominator $F(k)=0$,
\begin{equation}
F(k)=K \cot{(K-\delta_{\rm rs})}-\rmi k,
\end{equation}
defines a resonance point in the system. However, not every pole of $S$
is related with the compound system existence.

The resonance is usually observed in narrow region of energy~$E$,
which covers the resonant value $E_{\rm res}=E_0-\rmi \Gamma_0/2$
with some $E_0>0$ and $\Gamma_0>0$. Positivity of $E_0$ makes this
(quasi)discrete level unstable, whose lifetime is $\tau=2/\Gamma_0$.

In our model, we find $E_{\rm res}=K^2_{\rm res}-V$ for given $Q$, $V$,
and $\omega$ by solving the equation $F(\sqrt{K^2_{\rm res}-V})=0$
in an appropriate form. This equation is identically transformed and
is solved numerically, assuming $\omega^4\ll1$ and using
an iterative procedure with the initial value $K=\sqrt{V+Q}$:
\begin{eqnarray}
K^2_{m+1}&=&V+Q\nonumber\\
&&-\omega^4\left[\Delta(K_m)
+\rmi\gamma(K_m)\,\frac{k_m\cot{K_m}-\rmi K_m}{K_m\cot{K_m}-\rmi k_m}\right],
\nonumber\\
k_m&=&\sqrt{K^2_m-V}.\label{eqKK}
\end{eqnarray}
Omitting the indexes $m$ and $m+1$, one gets the equation
equivalent to $F(\sqrt{K^2-V})=0$.

To extract the resonance part of $S$-matrix, let us expand the complex
function $F(k)$ of a real $k$ in vicinity
of complex $k_{\rm res}=\sqrt{E_{\rm res}}=k_r-\rmi\kappa_r$ as
\begin{eqnarray}
F(k)&\simeq&F(k_{\rm res})+F^\prime(k_{\rm res}) (k-k_{\rm res})
\nonumber\\
&=&C\, (k-k_r+\rmi \kappa_r),
\end{eqnarray}
where $C=F^\prime(k_{\rm res})$ is a complex constant.

Since $k=k^*_{\rm res}$ solves conjugate equation $F^*(k)=0$,
and $F^*(k)\simeq C^* (k-k_r-\rmi \kappa_r)$ near the resonance,
we define the phase shift $\delta_0$ associated with potential
scattering so that
\begin{equation}
\rme^{2\rmi\delta_0}=\rme^{-2\rmi k}\,\frac{C^*}{C}.
\end{equation}

The calculated ingredients allow us to write down the scattering
matrix and the total phase shift in vicinity of the resonance:
\begin{eqnarray}
S&\simeq&\rme^{2\rmi\delta_0}\,\frac{k-k_r-\rmi\kappa_r}{k-k_r+\rmi\kappa_r},\\
\delta&=&\delta_0-\arctan{\frac{\kappa_r}{k-k_r}}.
\end{eqnarray}

Presenting these quantities in the conventional form, depended on $E$,
we expand $F(\sqrt{E})$ in powers of $E$:
\begin{equation}
F(\sqrt{E})\simeq \frac{C}{2k_{\rm res}}\,(E-E_{\rm res}),
\end{equation}
where $C$ is as above. It leads to redefinition
of phase shift~$\delta_0$ because of the relation:
\begin{equation}
\arctan{\frac{\kappa_r}{k-k_r}}-\frac{1}{2}\arctan{\frac{\kappa_r}{k_r}}
=\arctan{\frac{\Gamma_0}{2(E-E_0)}}.
\end{equation}

\begin{figure}
\begin{center}
\includegraphics[width=7.8cm]{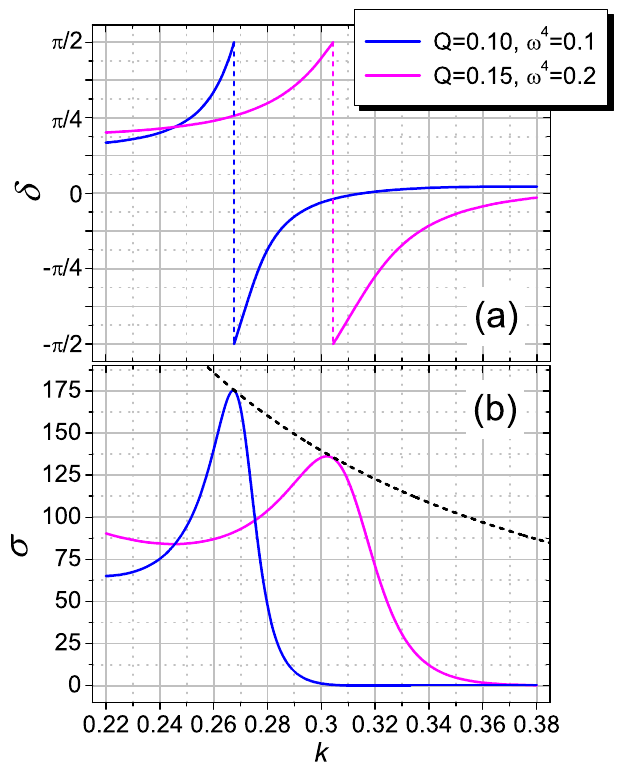}
\end{center}
\vspace*{-7mm}
\caption{Phase shift $\delta$ (a) and cross section $\sigma$ (b)
as functions of incident momentum $k$. Jumps and peaks in the graphs
indicate resonance and appear at $k_r={\rm Re}\,\sqrt{E_{\rm res}}$.
Black dashed line in (b) corresponds to maximal admissible value
$\sigma_{\rm max}=4\pi/k^2$.}\label{fig8} 
\end{figure}

Thus, we can see that the phase shift $\delta$ experiences
a jump $\delta(k_r-0)-\delta(k_r+0)=\pi$, which
reveals a resonance. Besides,
this determines the scattering cross section $\sigma$
in accordance with the optic theorem~\cite{LL}:
\begin{equation}
\sigma(k)=\frac{4\pi}{k^2} \sin^2{\delta(k)}.
\end{equation}

The behavior of $\delta(k)$ and $\sigma(k)$ (in dimensionless units)
is demonstrated in Fig.~\ref{fig8} and indicates that an excited two-particle
molecule can be created by an incident particle with a relative momentum
$k\in(k_r-\kappa_r/2;k_r+\kappa_r/2)$, where $k_r={\rm Re}\,\sqrt{E_{\rm res}}>0$
and $\kappa_r=-{\rm Im}\,\sqrt{E_{\rm res}}$.

The obtained formulas describe, in general, the resonance mechanism of
the short-term production of a two-particle molecule (dimer) without
specifying the interactions used. This means that the model still needs
to be refined in accordance with physical conditions, and our formalism
is applicable to the study of different scenarios. In principle, even
the energy gap $Q$ can be supported by a spatially homogeneous interaction
to create an energy predominance of one particle configuration over another.

In this connection, we mention experiments in the laboratory, where
the configurations depend on the spin, and the energy gap is associated
with the applied magnetic field~\cite{Grimm10}.
Adjustment of the gap $Q$ by a magnetic field leads to the Feshbach
phenomenon~\cite{Grimm10,BrH06}, that is a resonant transition
between configurations, accompanied by a jump in the scattering length
in open channel.

In the case of dark matter, until the interactions involved are clarified,
we minimize the number of interactions and demonstrate the capabilities of
our model.

For simplicity, let us put $V=0$, when $K=k$, and reduce Eq.~(\ref{eqKK})
to the form:
\begin{equation}\label{eqk2}
k^2_{m+1}=Q-\omega^4\left[\Delta(k_m)
+\rmi\gamma(k_m)\right],\quad
k_0=\sqrt{Q}.
\end{equation}
This case describes the transformation of a pair of asymptotically free
particles (scattered waves) into a dimer in the resonance zone due to
the potential $-12\cos_{\mu=1}{x}$ and the channel coupling~$\omega^2$.

Iterating Eq.~(\ref{eqk2}), we obtain the solutions:
\begin{equation}\label{E1}
E^{(11)}_{\rm res}=0.07317639856-\rmi\, 0.005600193629,
\end{equation}
for $Q_1=0.1$ and $\omega^4_1=0.1$; and
\begin{equation}\label{E2}
E^{(22)}_{\rm res}=0.09638881959-\rmi\, 0.01294943156,
\end{equation}
for $Q_2=0.15$ and $\omega^4_2=0.2$.

Having additionally calculated the resonance energies for the sets
of parameters $(Q_1,\omega^4_2)$ and $(Q_2,\omega^4_1)$:
\begin{eqnarray}
E^{(12)}_{\rm res}&=&0.04538086735-\rmi\, 0.008987063657,\label{E3}\\
E^{(21)}_{\rm res}&=&0.1236649388-\rmi\, 0.007215615748,\label{E4}
\end{eqnarray}
and comparing them with (\ref{E1}) and (\ref{E2}), we deduce that
increasing $\omega^4$ leads to an increase in the decay width of
the resonance.

We also note that the resonance energy $E_0={\rm Re}\,E_{\rm res}$
is less than the gap $Q$ for all cases at hand.

To complete the resonance description for two sets of parameters $Q$
and $\omega^4$, leading to (\ref{E1}) and (\ref{E2}), we obtain
numerically the potential phase shifts
$\delta^{(11)}_0(k;Q_1,\omega_1)\simeq-k+0.54331$
and $\delta^{(22)}_0(k;Q_2,\omega_2)\simeq-k+0.62674$, respectively.
The results of calculations are shown in Fig.~\ref{fig8}.

Since Eq.~(\ref{eqk2}) is identical to the condition $D(k)=0$
for the denominator of the wave functions at $V=0$, an incident
particle is capable to penetrate into the resonance zone at resonance
energy~$E_0$ and is actually reflected at other energies.

To convert dimensionless characteristics into physical units,
we use formulas (\ref{un2}). It can be seen that the scale $r_0$
is an essential parameter here, the determination of which
requires additional conditions that are beyond the scope of this
consideration. Although $r_0$ characterizes the interaction radius,
we constrain its value by the relativistic limit of a
non-relativistic model, noting that the Compton wavelength
$\hbar/(mc)$ is less than the de Broglie wavelength $\hbar/(mv)$
of slow particles at $v\to0$. Thus, replacing the reduced mass
in (\ref{un2}) with half the particle mass, we write the following
scales for distance, energy and time:
\begin{equation}\label{lim}
r_0=\frac{\hbar}{mc},\quad
\varepsilon_0=mc^2,\quad
\tau_0=\frac{\hbar}{\varepsilon_0},
\end{equation}
where the particle mass $m$ is used. Kinematically, $\tau_0=r_0/c$
is the time taken for a light to travel the distance $r_0$.
Besides, the wave number $k$ determines the momentum value $p=\hbar k/r_0=mck$.

For $\varepsilon_0\sim10^{-22}\,\text{eV}$, one has
\begin{eqnarray}
&&r_0\simeq1.973\times10^{15}\,\text{m}\,\left[\frac{mc^2}{10^{-22}\,\text{eV}}\right]^{-1},\\
&&\tau_0\simeq6.582\times10^6\,\text{s}\,\left[\frac{mc^2}{10^{-22}\,\text{eV}}\right]^{-1},
\label{tlim}
\end{eqnarray}
where $r_0$ is shorter than the typical scale of BEC DM (healing length $\sim10^{19}$~m)
in Sec.~\ref{S2} (see also \cite{GN21}). Although the shape of the dimer
wave function in Fig.~\ref{fig6} indicates the localization of one particle
near another, the influence between particles at large $r_0$ can be
maintained by an infinite scattering length of the long-range potential used.
Note that the interparticle distance in BEC DM is of the order
$(m/\rho)^{1/3}\sim10^{-13}$~m, where $\rho\sim10^{-20}\,\text{kg}\,\text{m}^{-3}$
is a mean mass density.

Defining the dimer lifetime as $t=2\tau_0/\Gamma_0$ and accounting for
$\tau_0\sim0.2$~year and $\Gamma_0\sim0.01$ as above, we deduce dimer
stability over a period $\sim40$~years. The long-term decay of the dimer
in this scenario suggests to treat DM as multicomponent because of
involving composites, whose presence affects also the BEC properties.
This fact stimulates a detailed study of aspects of the proposed
mechanism and the formation of complexes from several particles,
such as trimers.

\section{\label{S5}Concluding Remarks}

The main message of this work is to show that the axion-like self-interaction
as applied to ultralight DM bosons provides two phases (dilute and dense) of
the self-gravitating BEC DM, and its $\mu$-modification at the quantum mechanical
level ensures the appearance of composites, namely, dimers, using the Feshbach
resonance between two scattering channels. In our treatment, the creation of
individual complexes from initial particles with a large scattering length
precedes the BEC state and promotes its formation. This assumption can be used
to explain the very small scattering length (repeatedly confirmed by
calculations~\cite{Harko2011,Chavanis2,GN22}) of effective particles
in BEC, which then should be a mixture of composites. Such a scenario is 
induced by the experiments with atomic BEC~\cite{Grimm10,BrH06} and the resonant
transformations under strong interaction. In particular, the connection between
the observed intercepts of the $\pi$-meson correlations and the deformation parameter
$\mu$ used for their description is revealed in \cite{GM3} by accounting for
a quark structure and suggests a theoretical basis of describing composites
of the particles governed by various statistics.

The results shown here indicate the presence of two phases of the gravitating
BEC DM using the cosine-like self-interaction derived for QCD axions. This type
of interaction generalizes non-linear interactions in previous
studies~\cite{Harko2011,Chavanis2,GKN20,GN21} that help us interpret
innovations. In the present case, the existence of rarefied (gaseous) and
dense (liquid) states immediately follows from the initial conditions for
the Gross--Pitaevskii equation at zero temperature. A first-order phase
transition between these states should be stimulated by quantum fluctuations
that contribute to the pressure/compression, according to outcomes of
\cite{GKN20}. The influence of the two-phase structure on
the rotation curves of galaxies is expected to be similar to the results of
\cite{GN21}, but requires yet a detailed study. We admit that the contribution
of each phase depends on the formation stage of galaxy and DM halo.
However, as shown earlier in \cite{GN22}, the dense phase of BEC appears
unfavorable for composites of DM particles because of their probable destruction
caused by collisions.

To describe two-particle composites, in contrast to the heuristic approach
in \cite{GN22}, we turn to a quantum mechanical model with the smoothly
deformed cosine-like axion potential, which allows us to control the scattering
length changing the deformation parameter~$\mu\geq0$. Such potentials as functions
of the distance $r$ between two particles are studied for the first time
and turn out to be similar to typical atomic (molecular) potentials.
In other words, we have actually replaced the interaction of QCD axions
with an interaction inherent in atoms and molecules, and we argue 
the constructiveness of this change by the results obtained.

Thus, having got characteristics of $s$-wave potential scattering
in the Born approximation for the $\mu$-deformed potentials
$g\cos_{\mu}(r/r_0)$ with dimensionless coupling~$g$, we treat these
potentials as short-range ones for $\mu<1/2$, while the potentials
at $\mu>1/2$ behave as long-range ones with infinite scattering length.
It follows from the small-$k$ asymptotic $f\sim k^{\frac{1}{\mu}-2}$
of the scattering amplitude $f$ as function of wave number $k$ in
the linear approximation in the coupling constant $|g|<1$.
The strong coupling case requires the use of alternative
approach beyond the Born approximation~\cite{BrH06}.

Analyzing, we deduce that the problem with attractive potential for
$g=-12$ and $\mu=1$ is suitable for describing a dimer, that is,
a composite of two particles. In the case of $\mu=1$, one has
$\cos_{\mu}{x}\sim-x^{-2}$ at large~$x$ that allows one to associate
this problem to the Efimov's physics~\cite{BrH06}.
Although the general solution of the stationary Schr\"odinger
equation for fixed $\mu=1$, but arbitrary $g$ and some energy $E$ of
relative motion, is formulated in terms of the confluent Heun function,
the case of $g=-12$ and $E=0$ is unique because it leads to a particularly
simple solution for a dimer in the ground state. In this case, a quantum mean
squared displacement of moving particle in vicinity of $r=0$ is equal to
parameter $r_0$ of the potential used and corresponds to classically
admissible region. Strictly speaking, the range of $r_0$ is unknown,
and its value can be large enough, dealing with the potential of infinite
scattering length.

Our vision of forming such a dimer in space is based on the Feshbach
resonance and the use of two channels of two-particle scattering, when
a single bound state of dimer represents evidently a closed channel,
while the asymptotically free scattering of two particles corresponds
to an open channel. There is no tunneling within the one potential
problem here, but it rather appears a possibility to form an intermediate
compound system (dressed state), when a pair of particles jumps between
these channels with close energies by acting external perturbation.
Having positive energy of relative motion, the dressed state is
characterized by a finite lifetime and exhibits the resonance properties
to decay.

We have formulated a solution to the problem of two-channel scattering in
the first approximation, taking into account an additional potential
scattering in the open channel, which makes it possible to use the results
obtained to analyze various situations. In particular, the presence of
this type of scattering provides a nonvanishing resonance decay width
at zero energy, see (\ref{Rch}). Moreover, in the overall picture,
one would also have to take into account the possibility of resonant states
in the open channel~\cite{Marc04}. However, in an effort to minimize
the number of interactions between DM particles, we discarded the supplement
interaction in the open channel when obtaining the resonance energy $E_{\rm res}$
numerically. This means that apart from the interaction potential of the dimer,
we only need the coupling between the channels. In our approach, both the dimer
potential and the extra influence (\ref{exi}) are assumed to be characterized by
the same range $r_0$. Without additional research beyond the scope
of this work, it is impossible to estimate $r_0$, which determines
the resonance lifetime at the conventional particle mass
$m\sim10^{-22}~\text{eV}/c^2$. Limiting $r_0$ from above by the Compton
length, one obtains a long-lived dimer (about 40 years), see (\ref{tlim}).
Its fate depends on the properties of the open channel.
Anyway, the case of large $r_0$ may indicate its applicability
for describing larger structures of DM particles.

Thus, we are able to interpret dimers as constituents of DM, which
participate and contribute to forming BEC DM halo of galaxies. There
is a reasonings that the appearance of composites enhances the
formation of condensate~\cite{TBM}. Justifying this hypothesis
requires the use of quantum field and statistical approach to a
system of a large number of particles, which can serve as a prospect
for forthcoming research.

\acknowledgements

Both authors acknowledge support from the National Academy of Sciences of
Ukraine by its Project No.~0122U000888.


\end{document}